\documentclass[prd,email,preprint,showpacs,showkeys,preprintnumbers,amsmath,amssymb,nofootinbib]{revtex4}

\usepackage{amsmath}
\usepackage{latexsym}
\def\[{\left\lbrack}
\def\]{\right\rbrack}

\def\({\left(}
\def\){\right)}
\newcommand{\be}{\begin{equation}}
\newcommand{\ee}{\end{equation}}
\newcommand{\ea}{\end{eqnarray}}
\newcommand{\ba}{\begin{eqnarray}}

\newcommand{\p}{\partial}
\def\ni{\noindent}

\begin{document}

\pagestyle{myheadings}
\markright{N=2 supersymmetric radiation damping problem....}

%\topmargin 0mm
%\textwidth 175mm
%\textheight 240mm
%\evensidemargin 0mm
%\oddsidemargin 0mm
%\parskip=\medskipamount
%\def\baselinestretch{1.2}

\title{N=2 supersymmetric radiation damping problem \\ on a noncommutative plane}

%\author{E. M. C. Abreu$^a$\footnote{\sf E-mail: evertonabreu@ufrrj.br}, A. C. R. Mendes$^b$\footnote{\sf E-mail: albert@ufv.br}, C. Neves$^c$\footnote{\sf E-mail: clifford.neves@gmail.com} and W. Oliveira$^b$\footnote{\sf E-mail: wilson@fisica.ufjf.br}} 

\author{Everton M. C. Abreu$^{a,b}$} 
\email{evertonabreu@ufrrj.br}
\author{Albert C. R. Mendes$^b$}
\email{albert@fisica.ufjf.br}
\author{Wilson Oliveira$^b$} 
%\email{wilson@fisica.ufjf.br}
%\author{C. Neves$^d$}
%\email{cliffordneves@uerj.br} 

\affiliation{${}^{a}$Grupo de F\' isica Te\'orica e Matem\'atica F\' isica, Departamento de F\'{\i}sica, 
Universidade Federal Rural do Rio de Janeiro\\
BR 465-07, 23890-971, Serop\'edica, RJ, Brazil\\
${}^{b}$Departamento de F\'{\i}sica, ICE, Universidade Federal de Juiz de Fora,\\
36036-330, Juiz de Fora, MG, Brazil\\
%${}^{d}$Departamento de Matem\'atica e Computa\c{c}\~ao, Universidade do Estado do Rio de Janeiro\\
%Rodovia Presidente Dutra, km 298, 27537-000, Resende, Rio de Janeiro, Brazil\\
\bigskip
\today\\
{\bf Dedicated to the memory of Prof. Wilson Oliveira}}


\begin{abstract}
\ni It is well known that a direct Lagrangian description of radiation damping is still missing.  In this paper a specific approach of this problem was used, which is the standard way to treat the radiation damping problem.
A $N=2$ supersymmetric extension for the model describing the radiation damping on the noncommutative plane with electric and magnetic interactions was obtained. The entire supercharge algebra and the total Hamiltonian for the system were analyzed.  Finally, noncommutativity features were introduced and its consequences were explored..
\end{abstract}

\maketitle

\pagestyle{myheadings}
\markright{N=2 supersymmetric radiation damping problem.....}

\section{Introduction}
\label{intro}

%%%%%%%%%%%%%%%%%%%%%%%%%%%%%%%%%%%%%%%%%%%%%%%%%%%%%%%%%%%%%%%%%%%%%%%%%%%%%%%%%%%%%%%%%%%%%%%%%%%%%%%%%%%%%%%%%

%It is important to notice that it is unknown in the current literature a direct Lagrangian description of radiation damping (we will explain the difficulties in a moment).  It is not our intention here to present such a description.  We will work with a specific approach of this problem \cite{mt,bc,fv,cl2,cl3}.  This is the usual procedure treating the radiation damping issue.  Some new results were obtained.

%\bigskip
An underlying feature of all charged particles is that an accelerated charged particle radiate electromagnetic energy.  During this process, the recoil momentum of the emitted photons is equivalent to a reaction force relative to the self-interaction of the particle with its own electromagnetic field which creates the radiation damping \cite{hjp,becker,Lorentz}.

The analysis of dissipative systems in quantum theory has a great interest and relevance either because of fundamental reasons \cite{uz} or because of its applications in our real world \cite{cl,whb,walker,hk,Bula}.  
The explicit time dependence of the Lagrangian and Hamiltonian operators introduces a major difficulty in this study since the canonical commutation relations are not preserved in time.  
Different approaches have been used in order to apply the canonical quantization scheme to dissipative systems \cite{mt,bc,fv,cl2,cl3,galley}.

%One of these approaches focus on an isolated system composed by the original dissipative system plus a reservoir.  One start from the beginning with a Hamiltonian which describes the system, the bath and the system-bath interaction.  Subsequently, one eliminates the bath variables which give rise to both damping and fluctuations, thus obtaining the reduced density matrix \cite{cl,bc,fv,cl2,cl3}.

Another way to handle with the problem of quantum dissipative systems is to enlarge the target's phase-space in a way that we will have to deal with an effective isolated system composed by the original system plus its time-reversed copy \cite{ft,bm,bm2}.  The new degrees of freedom thus introduced may be  represented by a single equivalent (collective) degree of freedom for the bath, which absorbs the energy dissipated by the system.  
In order to implement a canonical quantization formalism, we must first double the dimension of the target phase-space.  The objective of this procedure is to obey the canonical quantization scheme, which requires an effective isolated system.

To study the quantum dynamics of an accelerated charge, we have to use indirect representations since the energy, the linear momentum and the angular momentum that are all carried by the radiation field are lost.  The consequences concerning the motion of the charge are known as radiation damping (RD) \cite{hjp}.

The reaction of a classical point charge to its own radiation was first discussed by Lorentz and Abraham more than a hundred years ago \cite{becker,Lorentz}.  There are two interesting aspects of the Abraham-Lorentz theory: the self-acceleration and pre-acceleration.  

Self-acceleration indicates the classical solutions, where the charge is under acceleration even in the absence of an external field.  Pre-acceleration means that the charge begins to accelerate before the force begins to act.

%We can say (again) that the main motivation to investigate the process of radiation damping is because it is important in many areas of the electron accelerator operation \cite{walker}.  It can be observed in recent experiments with intense-laser relativistic-electron scattering at laser frequencies and field strengths where radiation reaction forces begin to become significant \cite{hk,bula}.

A complete description of radiation damping is still missing.  Hence, in this paper we have discussed some aspects of RD framework concerning algebraic noncommutativity and supersymmetry, as well as the relative resulting physics, of course.  Notice that to talk about these issues in a RD system is very difficult because we have to deal with two systems (the particle and the reservoir) and consequently both mathematical and physical features are not trivial, as we will see.

In this work we have introduced a $N=2$ supersymmetric extension  for the RD model in addition to the $N=1$ supersymmetric version introduced in \cite{mendes,Barone}.  
%Also a new action dual equivalent to the RD one is obtained using the Noether dualization procedure.  Using the variables introduced in \cite{Plyushchay1,Plyushchay3} we obtain a new nonvanishing phase space for the Poisson brackets.  
We have used the nonrelativistic $(2+1)$-dimensional pseudo-Euclidean space model describing the RD (represented by the equation (\ref{01}) below) on the noncommutative (NC) plane which introduced an interaction term into the free model through the $N=2$ superfield technique.  
%To accomplish this task, the Lagrangian is  separated in two parts describing the ``external" and ``internal" degrees of freedom in a noncommutative phase space.   The introduction of a scalar superpotential for the interaction term makes it possible to construct the $N=2$ supersymmetric Lagrangian which allows us to analyze the interaction of our system with any kind of external potential (attractive or repulsive).  We used an harmonic-like potential in order to obtain equations that represents the damped (a general solution where the metric is general) and the standard (where the metric is the one used in this work) harmonic oscillators.  

As we said just above, it is important to notice that in fact there are two phase-spaces considered here.   The first one is where the RD occurs and the second one is the doubled phase-space where the details and the relevance will be described in section II and in the references quoted there.   It is in this doubled phase-space that we have carried out the considerations described in this work, i. e., noncommutativity and $N=2$ supersymmetry.  For example, concerning only the noncommutativity issue, it can be shown easily that the original space is commutative whereas in this doubled phase-space we will show precisely in this paper that it is NC.  We hope that our work can bring some light on the understanding of this extended space.

The organization of this paper is: in section 2 we will carry out a very brief review of the mechanical model with a Chern-Simons term developed in \cite{Lukierski} and its Galilean-symmetric version, i.e., the LSZ model.  
In section 3  we will present a symplectic structure for the model in order to introduce the noncommutativity through the variables used in \cite{Plyushchay1,Plyushchay3}.  In section 4 we will show the supersymmetric extension of the model, 
the supercharges and a supersymmetric version through the Hamiltonian formalism.  The conclusions and perspectives are described in the last section. 
 
\section{The model}

{\bf The LSZ model.} In \cite{Lukierski} the authors have introduced a nonrelativistic classical mechanics for the free particle model which is quasi-invariant under $D=2$ Galilei symmetry as
%the second term in (\ref{01.1}) was modified and we have that
\be \label{01.2}
L_{LSZ}=\frac{1}{2}\,m\,{\dot x_i}^2 \,-\,\kappa\varepsilon_{ij}\dot x_i \ddot x_j, \;\;i,j=1,2,
\ee
where $\kappa$ has dimensions of mass $\times$ time.  It can be shown [21] (by following the methods of ref. \cite{ls}) that this Lagrangian is quasi-invariant.  The model (\ref{01.2}) does not have a usual Galilei symmetry.  We can describe it by the exotic, two-fold centrally extended Galilei symmetry with non-commutating boosts. It was analyzed carefully in \cite{Lukierski} and later in \cite{duval2}.   
The authors in \cite{Lukierski} have demonstrated that the model describes the superposition of a free motion in an NC external and an oscillatory motion in an NC internal space.
A $N=2$ supersymmetric extension of (\ref{01.2}) was accomplished in \cite{Lukierski2} which analyzed particles in the NC space with electric and magnetic interactions.  A supersymmetrization of (\ref{01.2}) was firstly obtained in \cite{lsz}. 
%Both models above are depicted here to help the reader to understand the physical alternatives for actions like (\ref{01.1}) and (\ref{01.2}).  
Other considerations can be found in \cite{ah}.

%Recently, an action analogous to (\ref{01.1}) was constructed, although with another approach, in order to describe dissipative systems.

%\subsection{The radiation damping model}

\bigskip

{\bf The radiation damping model.}  In \cite{Barone,Albert1} another point of view concerning the study of RD \cite{hjp,hjp2,hjp3} was presented, where it was introduced a Lagrangian formalism in two dimensions given by
\be\label{01}
L_{RD}=\frac{1}{2}m\,g_{ij}\dot x_i \dot x_j -\frac{\gamma}{2}\varepsilon_{ij}\dot x_i \ddot x_j, \;\;i,j=1,2,
\ee
where $\varepsilon_{ij}$ is the Levi-Civita antisymmetric tensor and $g_{ij}$ is the metric for the pseudo-Euclidean plane \cite{bgpv} which is given by 
\be\label{02}
g^{ij}\,=\,g_{ij}=diag(1,-1)\,\,.
\ee
We are using the Einstein sum convention for repeated indices. The model (\ref{01}) was shown to have (1+1)-Galilean symmetry.  The dynamical group structure associated with the system is {\it SU}(1,1) \cite{Albert1}.  The supersymmetrization $N=1$ of (3) was studied in \cite{Albert2}. 

The Lagrangian (\ref{01}) describes, in this pseudo-Euclidean space, a dissipative system of a charge interacting with its own radiation, where the 2-system represents the reservoir or heat bath coupled to the 1-system \cite{mendes,Barone}.  It shows that the dissipative term, as a matter of fact, acts as a coupling term between both the 1-system and the 2-system in this space.  Specifically, we have a system formed by the charge and its time-reversed image, that globally behaves like a closed system described by equation (\ref{01}).

Notice that the Lagrangian (\ref{01}) is similar to the one discussed in \cite{Lukierski} (action (\ref{01.2})), which is a special nonrelativistic limit of  the particle with torsion \cite{Plyushchay}. 
However, in this case we have a pseudo-Euclidean metric and the RD constant ($\gamma$) which acts as a coupling constant of a Chern-Simons-like term. The RD constant $\gamma$ plays the same mathematical role of the ``exotic" parameter $\kappa$ in (2) \cite{Lukierski,Lukierski2}.  However, there is an underlying physical difference between both $\gamma$ and $\kappa$.

It is important to reinforce that the difference between the results that will be obtained here and the ones in \cite{lsz} is that, besides the metric, the physical systems are different.  The RD constant $\gamma$ is not a simple coupling constant.  It depends on the physical properties of the charged particle, like the charge $e$ and mass $m$ which are related to the objects in its equation of motion.  This last one depicts an interaction between the charge and its own radiation field.

\section{Noncommutativity}

The analysis of NC geometry and its applications in physics requires a great amount of time since it has many implications in several subjects such as quantum mechanics, high energy physics and condensed matter \cite{bcgms}.   As an example we can briefly describe the planar systems of condensed matter that deal with a perpendicular magnetic field and becomes itself NC in the lowest Landau level \cite{bcgms}.  Besides, the NC parameter can be identified with the inverse of the magnetic field.  The study of NC theories has received a special attention through the last years thanks to the possibility that noncommutativity can explain 
the physics of the Early Universe.  It has been used in many areas of theoretical physics \cite{nekrasov}, classical mechanics \cite{hms}, cosmology \cite{nosso} and Lorentz invariance \cite{amorim}. 

Let us introduce a Lagrangian multiplier which connects $\dot x$ to $z$ and after that we will substitute all differentiated $x$-variables in the Lagrangian 
(\ref{01}) by $z$-variables, one can construct a first-order Lagrangian
\be\label{03}
L^{(0)}\,=\,g_{ij} p_i (\dot x_j -z_j ) +\frac{m}{2}g_{ij} z_i z_j -\frac{\gamma}{2}\varepsilon_{ij}z_i \dot z_j\;\;,
\ee
where the equations of motion can be written through the symplectic structure \cite{Faddeev}, as
\be\label{04}
\omega_{ij}\dot \xi ^j =\frac{\partial H(\xi)}{\partial \xi^{i}}
\ee
and the symplectic two form is
\be\label{05}
(\omega)=
\begin{pmatrix}
{\bf0} & {\bf g} & {\bf0}\cr -{\bf g} & {\bf0} & {\bf0} \cr {\bf0} & {\bf0} & -\gamma \varepsilon 
\end{pmatrix}
\ee
where
\be\label{06}
\varepsilon=
\begin{pmatrix}
0 & 1 \cr -1 & 0 
\end{pmatrix},
\ee
and $g$ was given in Eq. (\ref{02}) and {\bf 0} denotes the $2\times\,2$ null matrix. $H(\xi^l)$ is the Hamiltonian and $\xi^i$ are the symplectic variables.

Let us use a modified version of the variables introduced in \cite{Plyushchay1,Plyushchay3} as
\ba\label{07}
{\cal Q}_i &=&\gamma\,g_{ij}(mz_j \,-\,p_j )\;,  \nonumber \\
X_i&=&x_i \,+\,\varepsilon_{ij}{\cal Q}_{j}\;, \nonumber \\
P_i&=&p_i\,\,.
\ea
We can consider that our Lagrangian can be divided in two disconnected parts in order to describe the ``external" and ``internal" degrees of freedom.  So, 
\be\label{08}
L^{(0)} =L^{(0)}_{\rm ext} + L^{(0)}_{\rm int}
\ee
where
\be\label{09}
L^{(0)}_{\rm ext}\,=\,g_{ij}\,P_i \dot X_j + {\gamma\over2}\varepsilon_{ij}P_i \dot P_j -{1\over 2m}g_{ij}P_i P_j ,
\ee
and
\be\label{10}
L^{(0)}_{\rm int}={1\over{2\gamma}}\varepsilon_{ij}{\cal Q}_i \dot{\cal Q}_j +{1\over{2m\gamma^2}}g_{ij}{\cal Q}_i {\cal Q}_j .
\ee

The internal coordinates, $\vec{{\cal Q}}$, and the external ones, $\vec{X}$, do not depend on each other \cite{Plyushchay1} and we can see that they do not commute.
The respective nonvanishing Poisson brackets are
\ba\label{11}
\{X_i ,X_j \} &=&\gamma \varepsilon_{ij},\;\;\; \{X_i ,P_j \}\,=\,g_{ij},\nonumber\\
\{{\cal Q}_i ,{\cal Q}_j \}&=& \gamma\varepsilon_{ij}\,\,,
\ea
where we can see clearly that the RD constant acts as the NC parameter that appears in the canonical NC approach.   Hence, we can conclude the presence of RD constant in any result from now on is a consequence of the NC effect. 
%To clarify, we can see in (\ref{08}) that our Lagrangian can be written as two separated and disconnected parts that describes both the ``external'' and ``internal'' degrees of freedom in a NC phase space, parameterized by the variables $(X_i ,P_i)$ (external structure) and ${\cal Q}_i$(internal structure) \cite{Plyushchay1,Plyushchay3}. 

Now we will introduce an interaction term into the ``external'' sector (equation (\ref{09})) which does not modify the internal sector and it is represented by a potential energy term $U(X)$ involving NC variables, as follows
\be\label{11.0}
L_{\rm ext}\,=\,g_{ij}\,P_i \dot X_j + {\gamma\over2}\varepsilon_{ij}P_i \dot P_j -{1\over 2m}g_{ij}P_i P_j -U(X)\,\,.
\ee
This result leads us to a deformation of the constraint algebra since the constraint now involves a derivative of the potential \cite{Albert1}.

%To end this section, notice that the $L^{(0)}$ has twice the phase-space dimension.  The ``external" and ``internal" dynamics should be interpreted in terms of underlying one-dimensional dissipative dynamics.  The same observation can be made for the results below.

Notice that the Lagrangian in Eqs. (10), (11) and (13) are formed by objects that show a NC algebra described in Eq. (12).  The standard NC procedure is to recover the commutative algebra in (12) using the so-called Bopp shift
\be \label{AXXXXX}
x_i \,=\, \hat{X}_i \,+\, \frac 12 \gamma \epsilon_{ij} p_j\,\,,
\ee
where the hat defines a NC variable.  Substituting (\ref{AXXXXX}) into (12) we will have that $\{x_i , x_j \} = 0$.  The same can be made with ${\cal Q}_i$ so that
\be \label{BXXXXX}
{\cal Q}_i\,=\,\hat{{\cal Q}}_i\,+\,\frac 12 \gamma \epsilon_{ij}  P_j\,\,,
\ee
where $\hat{p}_i\,=\,p_i$ and $P_i\,=\,\hat{P}_i$ and consequently $\{p_i , p_j \}=\{P_i , P_j\}=0$.   Substituting (\ref{AXXXXX}) and (\ref{BXXXXX}) into the Lagrangians (10), (11) and (13) results in a Lagrangian defined in NC phase-space.    

The NC effect of (8) is to separate the sectors, i.e., the external and internal ones.  Namely, the RD NC effect was to promote, from $D=2$ phase-space RD scenario, the analysis of $D=1$ dissipative dynamics.

\section{The $N=2$ supersymmetric model}

In \cite{luv} it was constructed the supersymmetric extension of the LSZ model in Eq. (2) in the NC plane.  An entire SUSY investigation was carried out and its $N=2$ extension was provided in \cite{Lukierski2} where interesting physical results were obtained.   The objective of this section is to construct the $N=2$ framework for RD in a NC plane since its $N=1$ was accomplished in \cite{Albert2}.   In this way our starting point is the Lagrangian in Eq. (13).   
%The reader will see that the results obtained here concerns an actual physical system, i.e., the RD , although the equations are analogous to the ones used in the $N=2$ analysis of LSZ action in (1).

To obtain the supersymmetric extension of the model described by the Lagrangian (\ref{11.0}), we will use a Grassmannian variable that will be connected with each commuting space coordinate which represents the system degrees of freedom. We are considering only $N=2$ SUSY for a non-relativistic particle, which is described by the introduction of two real Grassmannian variables $\Theta$ and $\bar\Theta$ (the Hermitian conjugate of $\Theta$) in the configuration space, but all the dynamics is parametrized by the time $t$ \cite{Galvão,Junker}.

Let us carry out the Taylor expansion for the real scalar supercoordinate as
\ba
\label{11.1}
&X_i& \rightarrow {\cal X}_i(t,\Theta,\bar\Theta)\,=\,X_i(t) + i\psi_i(t)\Theta + i\bar\Theta \bar{\psi}_i(t) +\bar{\Theta}\Theta F_i(t)
\ea
and their canonical supermomenta
\ba\label{11.2}
&P_i(t)& \rightarrow {\cal P}_i(t,\Theta,\bar\Theta)\,=\,i\eta_i(t) - i\Theta\left(P_i(t)+if_i(t)\right)-\bar{\Theta}\Theta\dot\eta_i(t),
\ea
which under the infinitesimal supersymmetry transformation laws
\ba\label{11.2.0}
\delta t=i\bar{\epsilon}\theta + i\bar{\epsilon}\Theta, \;\;\; \delta\Theta=\epsilon \;\;\;{\rm and}\;\;\; \delta\bar{\Theta}=\bar{\epsilon},
\ea
where $\epsilon$ is a complex Grassmannian parameter.  We can also write that 
\ba\label{11.2.1}
\delta {\cal X}_i &=&(\epsilon\bar{Q} + \bar{\epsilon}Q){\cal X}_i \\
\mbox{and} \qquad \delta{\cal P}_i &=& (\epsilon\bar{Q} + \bar{\epsilon}Q){\cal P}_i\,\,,
\ea
where both $Q$ and $\bar{Q}$ are the two SUSY generators
\be\label{11.2.3}
Q=\frac{\partial}{\partial \bar{\Theta}}+i\Theta\frac{\partial}{\partial t},\;\;\;\;\; \bar{Q}=-\frac{\partial}{\partial {\Theta}}-i\bar{\Theta}\frac{\partial}{\partial t}.
\ee

In terms of $(X_i(t), P_i(t), F_i,f_i)$, the bosonic (even) components and $(\psi_i(t),\bar{\psi}_i(t),\eta_i(t))$, the fermionic (odd) components, we can obtain the following supersymmetric transformations,
\ba\label{11.2.4}
\delta X_i \,=\, i(\bar{\epsilon}\bar{\psi}_i +\epsilon\psi_i ) &;&\quad 
\delta \psi_i \,=\, -\bar{\epsilon}(\dot{X}_i -iF_i )\nonumber\\
\delta \bar{\psi}_i \,=\, -\epsilon (\dot{X}_i +iF_i ) &;&\quad  
\delta F_i \,=\, \epsilon\dot{\psi}_i - \bar{\epsilon}\dot{\bar{\psi}}_i \,,
\ea
and
\be\label{11.2.5}
\delta\eta_i = \epsilon(P_i +if_i) ;\quad
\delta P_i = 0 ;\quad
\delta f_i = 2\bar{\epsilon}\dot{\eta}_i\;\;. 
\ee 
%
%Notice that the supersymmetry, obviously, mixes the even and odd coordinates. Carrying out a variation in the even components we obtain the odd components and vice-versa.

The super-Lagrangian for the super point particle with $N=2$, which is invariant under the transformations (\ref{11.2.4}) and (\ref{11.2.5}),  can be written as the following integral (we have used for simplicity that $m=1$)
\ba\label{11.3}
\bar{L}_{\rm ext}&=&\frac{1}{2}\int{\rm d}\Theta {\rm d}{\bar\Theta}\left[ \frac{}{}g_{ij}\(\bar{D}{\cal X}_i \bar{\cal P}_j+{\cal P}_jD{\cal X}_i\)\right. \nonumber\\ 
&+&\left. \frac{\gamma}{2}\varepsilon_{ij}\({\cal P}_i\,{\dot{\bar{\cal P}}}_j \,+\,{\dot{\cal P}}_j \,{\bar{\cal P}}_i \) -\frac{1}{2}g_{ij}\({\cal P}_i\bar{\cal P}_j +{\cal P}_j\bar{\cal P}_i \)\right]\nonumber\\
&-& \int{\rm d}\,\Theta {\rm d}\,{\bar\Theta}U[{\cal X}(t,\Theta,\bar\Theta )]
\ea
where $D$ is the covariant derivative $(D=\partial_{\Theta}-i\bar\Theta\partial_t)$ and $\bar D$ is its Hermitian conjugate. The superpotential $U[{\cal X}]$ is a polynomial function of the supercoordinate

Let us expand  the superpotential $U[{\cal X}]$ in Taylor series and if we maintain $\Theta\bar\Theta$ (because only these terms survive after  integrations over Grassmannian  variables $\Theta$ and $\bar\Theta$), we have that
\ba\label{11.4}
U[{\cal X}]&=&{\cal X}_i \frac{\partial U[X(t)]}{\partial X_i} + \frac{{\cal X}_i{\cal X}_j^*}{2}\frac{\partial^2 U[X(t)]}{\partial X_i \partial X_j}+ ...\\
&=&F_i \bar\Theta \Theta \partial_i U[ X(t)]+\bar\Theta \Theta \psi_i \bar{\psi}_j \partial_i \partial_j U[ X (t)]+...\nonumber
\ea
where the derivatives $\partial_i =\frac{\partial}{\partial X_i}$ are such that $\Theta=0=\bar\Theta$, which are functions only of the $X(t)$ even coordinate. Substituting equation (\ref{11.4}) into equation (\ref{11.3}), we can write, after integrations that
\ba\label{11.5}
\bar{L}_{\rm ext}&=& L^{(0)}_{\rm ext} -\frac{1}{2}g_{ij}f_i f_j-g_{ij}F_i f_j +\frac{\gamma}{2}\varepsilon_{ij}f_i\dot{f}_j \nonumber\\ &-&big_{ij}\(\bar{\psi}_i \dot{\bar \eta}_j  -\dot{\eta}_j\psi_i \) -big_{ij}\dot{\eta}_i \bar{\eta}_j +i\gamma \varepsilon_{ij}\dot{\eta}_i\dot{\bar\eta}_j \nonumber\\&-&F_i \partial_i U[X(t)]-\psi_i \bar{\psi}_j \partial_i \partial_j U[X (t)],
\ea
which is the complete Lagrangian for $N=2$. 

The bosonic component $F_i$ is not a  dynamic variable. In this case, using the Euler-Lagrange equations for the auxiliary variables $f_i$ and $F_i$, we obtain that
\ba
f_i(t)&=& g_{ij}\partial_j U[X(t)],\label{11.6}\\
F_i(t)&=& f_i +\gamma g_{il}\varepsilon_{lj}\dot f_j \nonumber\\
&=&g_{ij} \partial_j U[X(t)]-\gamma\varepsilon_{ij} \partial_j\partial_k U[X] \dot X_k (t),\label{11.7}
\ea
where we have to eliminate the variable $f_i$ as well as its derivative in $F_i$.  Now, if we substitute (\ref{11.6}) and (\ref{11.7}) into  (\ref{11.5}) the auxiliary variables can be completely eliminated, hence 
\ba\label{11.8}
\bar{L}_{(N=2)\rm{ext}}&=&L_{\rm{ext}}^{(0)} -\frac{1}{2}g_{ij}\partial_i U\partial_j U+\frac{\gamma}{2}\varepsilon_{ij}\partial_i U\partial_j\partial_k U\dot X_k \nonumber \\&-&big_{ij}\(\bar{\psi}_i \dot{\bar \eta}_j  -\dot{\eta}_j\psi_i \) -big_{ij}\dot{\eta}_i \bar{\eta}_j + i\gamma \varepsilon_{ij}\dot{\eta}_i\dot{\bar\eta}_j\nonumber\\ &-&\psi_i\bar{\psi}_j \partial_i\partial_j U\,\,.
\ea

Note that, as in \cite{Lukierski2}, we can rewrite equation (\ref{11.8})
 as 
\ba\label{11.9}
\bar{L}_{(N=2)\rm{ext}}&=&{L}_{\rm{ext}}^{(0)} +A_k(X,t)\dot X_k +A_0(X,t)+ \nonumber \\&-&big_{ij}\(\bar{\psi}_i \dot{\bar \eta}_j  -\dot{\eta}_j\psi_i \) -big_{ij}\dot{\eta}_i \bar{\eta}_j + i\gamma \varepsilon_{ij}\dot{\eta}_i\dot{\bar\eta}_j\nonumber\\ &-&\psi_i\bar{\psi}_j\partial_i\partial_j U,
\ea
which is invariant under standard gauge transformations $A_{\mu}\rightarrow A_{\mu}^{\prime}=A_{\mu} +\partial_{\mu}\Lambda$, where
\be\label{11.10}
A_0( X,t) =-\frac{1}{2}g_{ij}\partial_i U\partial_j U
\ee
and 
\be\label{11.12}
A_k(X,t)=\frac{\gamma}{2}\varepsilon_{ij}\partial_i U\partial_j\partial_k U\,\,,
\ee
where both can be identified in \cite{Lukierski2} with the scalar potential $A_0$ (in this case we have a pseudo-Euclidean metric) and the vector potential $A_k$.  
Notice that both potentials above are not independent.
The vector potential introduces a magnetic field $B=\varepsilon_{ij}\partial_i A_j$ given by
\be\label{11.13}
B( X)=\frac{\gamma}{2}\varepsilon_{il}\varepsilon_{jk}\left(\partial_i\partial_l U\right)\left(\partial_j\partial_k U\right)\,\,,
\ee
where we can see that the noncommutativity introduced by the parameter $\gamma$ generates both a constant magnetic field \cite{Lukierski2} and an electric field given by $E_i\,=\,\p_i\,A_0$ which can be written as
\be\label{11.13.1}
E_i\,(X)\,=\,-\,g_{jk}\,\p_i\,\p_j\,U\,\p_k\,U\,\,.
\ee

The Euler-Lagrange equations, in this case, are
\begin{subequations}
\label{11.14}
\ba
m^* \dot X_i \,=\, P_i \,&-&\,me\gamma\varepsilon_{ij}E_j \,+\,\, m\gamma\varepsilon_{ij}\psi_l\bar{\psi}_k\partial_l\partial_k\partial_j U, \label{11.14a}\\
\dot P_i \,=\, e\,g_{ij}\varepsilon_{jl}\dot X_{l}B &+& beg_{ij}E_j \,-\, g_{ij}\psi_l\bar{\psi}_k\partial_l\partial_k\partial_j U,\label{11.14b}
\ea
\end{subequations}
where $E_i$ and $B$ are the electric and magnetic field, respectively, and $$m^{*} =m(1-e\gamma B)$$ is an effective mass. 
Notice that noncommutativity introduces a correction term in order to obtain an effective mass for the system.

However, this way of introducing electromagnetic interaction modifies the symplectic structure of the system which determines the NC phase-space geometry, for the bosonic sector, equation (\ref{11}), we have
\ba\label{11.15}
\{X_i,X_j \}&=&\frac{m}{m^*}\gamma\varepsilon_{ij},\;\; \{X_i ,P_j \}=\frac{m}{m^*}g_{ij},\nonumber\\
\{P_i ,P_j \}&=&\frac{m}{m^*}b\varepsilon_{ij},
\ea
where we have the value $e\gamma B \neq 1$ in order to avoid a singularity \cite{Plyushchay2,duval}.  Notice that the algebra in (\ref{11.15}) is different from the one in (12) where the momenta commute. In other words, the noncommutativity introduces a new NC algebra to the system besides the modification in the old one.

Concerning the fermionic sector, the Euler-Lagrange equations are
\ba\label{11.16}
i\gamma\varepsilon_{ij}{\ddot{\bar\eta}}_j +big_{ij}{\dot{\bar\eta}}_j -i\dot\psi_i &=&0,\nonumber\\
i\gamma\varepsilon_{ij}{\ddot{\eta}}_j +big_{ij}{\dot\eta}_j -i\dot{\bar\psi_i} &=&0,
\ea
for the fermionic variables $(\eta,\bar\eta)$. For the fermionic variables  $(\psi_i ,\bar{\psi}_i )$ the Euler-Lagrange equations are
\ba\label{11.17}
i\dot\eta_i + g_{ik}\bar\psi_j \partial_k\partial_j U &=& 0\nonumber\\
i\dot{\bar\eta}_i - g_{ik}\psi_j \partial_j \partial_k U &=&0\,\,.
\ea
where the fermionic variables $(\psi_i, \bar{\psi}_i )$ do not have dynamics.

So, analogously to \cite{Lukierski2} we have here that noncommutativity have introduced electric and magnetic fields into the system.  In the case of RD, studied here, described in an NC hyperbolic phase-space, the movement of a charged particle have an extra electromagnetic energy that did not appear in an $N=1$ SUSY analysis \cite{Albert2}.  This result agrees with the fact that noncommutativity does not change the physics of the system.  However, we understand that this electromagnetic energy is an extra one due to the NC feature of the phase-space.  This result is also different, as it should be expected, from the one obtained in \cite{Lukierski2} where only a magnetic interaction appear.

\subsection{The harmonic oscillator solutions}

In order to obtain an interesting solution of equations (\ref{11.14}) let us consider a specific form for the superpotential
\be\label{446}
U(X)\,=\,\frac{\omega}{2}\,g_{ij}\,X_i\,X_j\,\,,
\ee
which has clearly an harmonic-like form.

It is easy to see that in both equations (\ref{11.14a}) and (\ref{11.14b}) the last term that have three derivatives disappear and so we have two new equations that show the separation of both the fermionic and bosonic sectors, so that,
\begin{subequations}
\label{447}
\ba
& &m^*\,\dot{X}_i \,=\, P_i\,-\,e\,\gamma\,\epsilon_{ij}\,E_j \label{447a}\\
& &\!\!\!\!\!\!\!\!\!\!\!\!\!\!\!\!\!\!\!\!\!\!\!\!\!\!\!\!\!\!\!\!\!\!\!\!\!\!\!\!\!\!\!\!\!\!\!\!\!\!\!\!\!\!\!\!\!\!\!\!\!\!\!\!\!\!\!\!\!\!\!\!\!\!\!\!\!\!\!\!\!\!\!\mbox{and} \nonumber \\
& &\dot{P}_i \,=\, e\,g_{jl}\,\epsilon_{ij}\,\dot{X}_l\,B\,+\,e\,g_{ij}\,E_j \label{447b}\,\,.
\ea
\end{subequations}

\ni Computing a second time derivative of equation (\ref{447a}) and using (40b) we have that
%\be\label{448}
%m^*\,\ddot{X}_i\,=\,\dot{P}_i\,-\,e\,\gamma\,\epsilon_{ij}\,\dot{E}_j\,\,.
%\ee

%\ni Substituting (\ref{447b}) into (\ref{448}) we have that
\be\label{449}
m^*\,\ddot{X}_i\,=\,  e\,g_{ij}\,\epsilon_{jl}\,\dot{X}_l\,B\,+\,e\,g_{ij}\,E_j \,-\,e\,\gamma\,\epsilon_{ij}\,\dot{E}_j\,\,
\ee

\ni and this differential equation will disclose a very well known result.

From (\ref{11.10}), (\ref{11.12}) and (\ref{446}), we can write that,
\begin{subequations}
\label{450}
\ba
A_0\,(X,t) &=& -\,\frac{\omega^2}{2}\,g_{ij}\,X_i\,X_j \,\,,\label{450a}\\
A_k\,(X,t) &=& \frac{\gamma}{2}\,\omega^2\,\epsilon_{kj}\,X_j\,\,.\label{450b}
\ea
\end{subequations}

\ni Substituting these equations into (\ref{11.13}) and (\ref{11.13.1}) we have that
\be\label{551}
B\,=\,\gamma\,\omega^2
\ee
and
\be\label{552}
E_i\,=\,-\,\omega^2\,g_{ij}\,X_j\,\,,
\ee

\ni and finally, substituting these both equations into (\ref{449}) it is easy to show that
%\begin{wide text}
\be\label{553}
\ddot{X}_i\,-\,\frac{\gamma\,e\,\omega^2}{1-\gamma^2\,\omega^2\,e}\,(\,g_{ij}\,\epsilon_{jk}\,+\,g_{jk}\,\epsilon_{ij}\,)\,\dot{X}_k 
\,+\,\frac{e\,\omega^2}{1-\gamma^2\,\omega^2\,e}\,X_i\,=\,0\,\,,
\ee
%\end{wide text}

\ni which is the equation of a damped harmonic oscillator.   We can see clearly that the second-term of (\ref{553}) represents a dissipative force proportional to the velocity and in the last term of (\ref{553}), we have that
$$\omega_0^2\,=\,\frac{e\,\omega^2}{1-\gamma^2\,\omega^2\,e}\,\,$$

\ni can be seen as the natural frequency of this oscillator, $\omega_0$.  Notice that the RD constant is responsible for the dissipative force and it also affects the frequency.    The instantaneous rate of energy of the oscillator in (\ref{553}) can be written as
\be\label{554}
\frac{dE}{dt}\,=\,m^*\,\frac{\gamma\,e\,\omega^2}{1-\gamma^2\,\omega^2\,e}\,\dot{X}_i\,\,,
\ee
so that the RD constant also affects the energy rate.  Notice that when $\gamma=0$ we have the standard and well known results.  We have also that $\gamma^2 \omega^2 e \not= 1$, obviously.
If we have the metric in (63) and we use the pseudo-Euclidean plane given in (4), we can see that the second term in (\ref{553}) disappears and we have that
$$\ddot{X}_i\,+\,\frac{e\,\omega^2}{1-\gamma^2\,\omega^2\,e}\,X_i\,=\,0$$
$$\Longrightarrow \ddot{X}_i\,+\,\omega_0^2\,X_i\,=\,0\,\,,$$

\ni which is the equation for the standard harmonic oscillator that has the standard solutions.  However, the difference is the NC contribution.

From (\ref{553}) we can see that, since there is not a term which has three derivatives of $X$, one can conclude that in the NC space, the non-physical solutions, namely the pre-acceleration solutions (for $\dddot{X}$), do not exist.

\subsection{The supercharge algebra}

Now, from the supersymmetric transformations, equations (\ref{11.2.4}) and (\ref{11.2.5}), and the Lagrangian (\ref{11.9}), we can compute the supercharge algebra, through the Noether's theorem. The results for the charge operator are given by
\be\label{11.17.1}
Q=ig_{ij}(P_i -iW_i )\psi_j \;\;\; {\rm and} \;\;\; \bar{Q}=ig_{ij}(P_i + iW_i)\bar{\psi}_j \;\;,
\ee
where $W_i (X) =\partial_i U(X)$.

The supercharge algebra is
\be\label{11.7.2}
\{Q,Q\}=\{\bar{Q},\bar{Q}\} =0\,\,,
\ee
and
\be\label{11.7.3}
\{Q,\bar{Q}\} =-2i H .
\ee
Moreover, we can easily carry out a canonical calculation of the Hamiltonian and we obtain that
\be\label{11.7.4}
H=H_{b} + H_{f},
\ee
where the bosonic Hamiltonian $H_{b}$ is given by
\be\label{11.7.5}
H_{b}=\frac{1}{2}g_{ij} \(P_i P_j + W_i W_j \)\; ,
\ee
and the fermionic part $H_f$ can be written as
\be\label{11.7.6}
H_f = \frac{m}{m^{*}}\Big[ ieB(X)\varepsilon_{ij} \bar{\psi}_i \psi_j +g_{ik}\partial_j W_k (X) \bar{\psi}_i \psi_j \Big].
\ee
Note that the second term in $H_b$ is proportional to the scalar potential, equation (\ref{11.10}), i. e., there is a potential energy term in $H_b$.  We can say that the origin of this term is related to the electric field.

There is an alternative way to introduce the minimal electromagnetic interaction.  It can be accomplished through the transformation $P_i \rightarrow {\cal P}_i =P_i + eA_i (X_i ,t)$ in the Hamiltonian that keeps the symplectic structure of equation (\ref{11}). In \cite{Lukierski2} this transformation has been considered and the authors have obtained the same expression for the magnetic field equation (\ref{11.13}).

\section{Remarks and conclusions}

A fundamental property of all charged particles is that the electromagnetic energy is radiated whenever they are accelerated.  The recoil momentum of the photons emitted during this process is equivalent to a reaction force corresponding to the particle self-interaction with its own  electromagnetic field, which originates the RD.

%The process of RD is important in many areas of the electron accelerator operation, like in recent experiments with intense laser relativistic electron scattering at laser frequencies and field strengths where radiation reaction forces begin to become significant.
 
%In \cite{Albert1} some of us introduced an alternative approach to canonical quantization of the RD relying on doubling the degrees of freedom.  A Lagrangian model for the system with a Chern-Simons-like term with high order derivative was obtained (equation (3)).   In \cite{Albert2}  it was introduced the $N=1$ supersymmetric version of the RD in the Grass man super space.

Here the supersymmetric model was separated in two parts, namely, ``external" and ``internal" degrees of freedom of the supersymmetric model in terms of the new variables, where the RD constant introduced noncommutativity into the coordinate sector.   We have shown a way to introduce an electromagnetic coupling.  

We have calculated the supersymmetric $N=2$ extension of the RD model.   We have demonstrated that the noncommutativity introduced by the parameter generated a constant magnetic field.  With this result, combined with the electric field we have obtained a general expression for the damped harmonic oscillator which results in the standard harmonic oscillator in our pseudo-Euclidean space.  We have seen that in the NC space, the non-physical solutions, namely the pre-acceleration solutions, have disappeared.  After that, we have computed the supercharges algebra and the total Hamiltonian of the system, which was divided into two parts: bosonic and fermionic.

%Besides, we have used an alternative way to construct a dual equivalent action to the RD one and we have used the Noether dualization procedure.  We showed that the RD action is self-dual and also that, despite the LSZ action can be transformed in the RD one, both have different symmetries.  The dualization procedure showed precisely that although both actions are mathematically equivalent, they are physically different thanks to its coupling constant.  The RD coupling constant depends on each problem while for the LSZ action, it is simply a constant parameter.   Although it can sound like an obvious thing, however it is not.
%It used the Noether technique which is independent of dimensions and imposes a gauge symmetry which is believed to be hidden in the theory.  The main ingredient is an auxiliary field which is eliminated through the equations of motion and the final action is an effective one depending only on this original variables.

%With this new features revealed here we hope that this work has improved the comprehension of this extended space, which we believe it is not entirely well understood in the current literature.  

A perspective for future analysis is to study some typical problems of dissipative systems, like self-acceleration and pre-acceleration, for example.   To accomplish this, in our $N=2$ supersymmetric case, we have to analyze the Euler-Lagrange equations (\ref{11.14}), (\ref{11.16}) and (\ref{11.17}).

\section{Acknowledgments}

\ni The authors would like to thank CNPq (Conselho Nacional de Desenvolvimento Cient\' ifico e Tecnol\'ogico) and FAPEMIG (Funda\c{c}\~ao de Amparo \`a Pesquisa do 
Estado de Minas Gerais), Brazilian scientific support agencies, for partial financial support.

\end{document}